\documentclass[10pt]{article}
\usepackage{graphicx, xcolor}
\usepackage{amsmath, amssymb, amsthm}
\usepackage[margin=1in]{geometry}
\usepackage{booktabs} 
\usepackage{enumerate} 
\usepackage{enumitem}
\usepackage[colorlinks=true,linkcolor=blue,citecolor=blue]{hyperref}
\usepackage{xspace}
\RequirePackage[utf8]{inputenc}
\RequirePackage{apacite}

\title{Uncovering Student Engagement Patterns in Moodle with Interpretable Machine Learning}
\author{{\large Laura J. Johnston}\\Department of Statistical Science, UCL\\laura.johnston.22@ucl.ac.uk \and {\large Jim E. Griffin}\\Department of Statistical Science, UCL\\j.griffin@ucl.ac.uk  \and {\large Ioanna Manolopoulou}\\Department of Statistical Science, UCL\\i.manolopoulou@ucl.ac.uk \and {\large Takoua Jendoubi}\\Department of Statistical Science, UCL\\t.jendoubi@ucl.ac.uk }
\date{}

\begin{document}

\flushbottom 

\maketitle 


\begin{abstract}
Understanding and enhancing student engagement through digital platforms is critical in higher education. This study introduces a methodology for quantifying engagement across an entire module using virtual learning environment (VLE) activity log data. Using study session frequency, immediacy, and diversity, we create a cumulative engagement metric and model it against weekly VLE interactions with resources to identify critical periods and resources predictive of student engagement.

In a case study of a computing module at University College London's Department of Statistical Science, we further examine how delivery methods (online, hybrid, in-person) impact student behaviour. Across nine regression models, we validate the consistency of the random forest model and highlight the interpretive strengths of generalised additive models for analysing engagement patterns.

Results show weekly VLE clicks as reliable engagement predictors, with early weeks and the first assessment period being key. However, the impact of delivery methods on engagement is inconclusive due to inconsistencies across models. These findings support early intervention strategies to assist students at risk of disengagement. This work contributes to learning analytics research by proposing a refined VLE-based engagement metric and advancing data-driven teaching strategies in higher education.\\ 

{\parindent0pt
\textbf{Keywords:} Learning analytics, educational data mining, machine learning, generalised additive model, random forest, student engagement, virtual learning environment
}
\end{abstract}




\section{Introduction}
\addcontentsline{toc}{section}{Introduction} 
\label{sec:intro}
Student engagement, encompassing cognitive, emotional, and behavioural investment in learning \cite{reeve&tseng2011,appletonetal2008}, is integral to educational success. It not only predicts academic achievement and student well-being \cite{klem&connell2009,sinatraetal2015} but also contributes to the vitality of educational institutions. Understanding and fostering student engagement across varied contexts is essential given the diversification of education delivery modes, including in-person and online formats. 
 
\subsection{Challenges in Quantifying Student Engagement}
 
Quantifying student engagement is complex due to its multifaceted nature. Traditional methods include self-reported surveys \cite{greene2015,appleton2006} and observational techniques \cite{lane&harris2015,renninger&bachrach2015}, which can be influenced by subjective biases. Objective measures such as attendance and online activity hours \cite{rodgers2008,veettil2020} and academic performance often capture only specific aspects of engagement, varying among individuals and contexts. Moreover, the shift towards varied educational delivery methods, significantly accelerated by the COVID-19 pandemic, introduces additional layers of complexity in measuring engagement. While some studies have begun to explore these dynamics \cite{abdelkaderetal2020,westeretal2021,dascaluetal2021}, there remains a significant gap in models that account for the nuances introduced by online, hybrid, and in-person delivery modes.
 
The integration of educational data mining (EDM) and learning analytics (LA) with virtual learning environments (VLEs), such as Moodle, offers new avenues for measuring student engagement. VLEs generate substantial data, including activity logs that record student interactions—from page views and content access to discussions and assignment submissions. \citeA{pardo2014} suggest that log data can effectively monitor online student engagement, while \citeA{motzetal2019} found correlations between VLE log-derived engagement levels and those perceived by lecturers. However, critics argue that activity logs fail to capture all relevant information, such as the depth of engagement, the quality of learning \cite{wong&chong2018} or all aspects of behavioural, emotional, and cognitive student engagement \cite{kimetal2023}. 
 
Recent studies have incorporated multiple indicators from VLE data to quantify engagement \cite{caspari2022}. \citeA{golchehreh2023} categorised these indicators into three themes: VLE login and usage \cite{kitturetal2022,froissardetal2015,boteloenzoetal2017}, academic performance via assignments and assessments \cite{akpinaretal2020}, and communication through messaging and forums \cite{gorgunetal2022}. These studies quantify student engagement and subsequently apply modelling techniques to predict engagement-related behaviours, such as academic performance or dropout rates. This methodological approach quantifies engagement and validates its effectiveness and accuracy by assessing how well these engagement metrics can predict key student outcomes. 
 
\subsection{Modelling Approaches for Student Engagement}
 
Studies have employed various statistical and machine learning models to analyse VLE data. Random forest regressors and classifiers have shown strong performance with this data type \cite{gorgunetal2022,bulathwela2020,sivaneasharajah2020,akpinaretal2020,zhangetal2021}, capturing complex relationships and offering some interpretability through variable importance scores. Regression models, such as ridge regression \cite{bulathwela2020,zhangetal2021}, lasso \cite{yamasari2021}, and elastic net regression \cite{huetal2017}, are preferred when interpretability is paramount due to their straightforward insights into predictor effects. Models like support vector regression (SVR), neural networks, and deep learning \cite{dascaluetal2021,jiangandbosch2022,ongetal2022,wangetal2020} perform well with nonlinear data but often function as “black boxes,” lacking interpretability.

A notable gap in the current literature is the limited use of generalised additive models (GAMs) \cite{hastie1986}. GAMs bridge the gap between flexibility and interpretability by capturing complex, nonlinear relationships while maintaining a high degree of interpretability through smooth, additive functions for each predictor. This allows researchers to understand and visualise the individual effect of each predictor on the outcome variable. Consequently, GAMs offer superior interpretability compared to other nonlinear models, making them particularly well-suited for educational data, where understanding the influence of specific engagement indicators on student outcomes is as critical as accurately modelling them.

Feature engineering from VLE activity logs often involves tracking specific student actions or interactions with learning materials by counting activities over defined periods, such as days, weeks, or study sessions 
\cite{sisovic2015,app10010354}. This method includes metrics like the number of times students interact with resources—such as watching or pausing videos 
\cite{nagrecha2017}, submitting quizzes 
\cite{chen2020,sebaeetal2019}, total time spent on VLE activities 
\cite{kitturetal2022}, and counts of study sessions and page views 
\cite{kloftetal2014}. Similarly, 
\citeA{boteloenzoetal2017} defined engagement based on interactions with primary resources for each chapter. This approach easily adapts to different courses and materials and provides a straightforward measure of student behaviour. 

\subsection{Our Approach}
 
Building upon existing methods, we emphasise student engagement as the primary outcome, leveraging its observable and quantifiable nature. We propose a novel method for quantifying student engagement, extending the metric developed by \citeA{wong&chong2018}. This method aggregates multiple dimensions—frequency, immediacy, and diversity of student study sessions across chapters—into a singular value for the academic year. The metric aligns with themes identified by \citeA{golchehreh2023}; frequency and immediacy measure VLE logins and usage, while diversity captures quizzes, assignments, and forum participation. 
 
While this metric comprehensively captures engagement dimensions, its reliance on year-long data limits its utility for real-time interventions. We focus on weekly interactions as immediate engagement indicators to address this, offering a dynamic and actionable approach. Our model aims to correlate these short-term behaviours with overall engagement, facilitating early pattern detection and timely educational interventions. Additionally, we include the module’s delivery method (online, hybrid, or in-person) as a variable to account for its potential influence on student engagement.
 
Our study addresses the following research questions: 
 
\begin{description} 
\item[RQ1] How can student engagement be quantified using VLE data? 
\item[RQ2] How can student engagement be modelled using weekly interactions with resources as predictors? 
\item[RQ3] How can we interpret which weeks and resources are most important in predicting student engagement while accounting for the module’s delivery method?
\end{description} 

To address \textbf{RQ1}, we refine the engagement metric proposed by \citeA{wong&chong2018}, adapting it to modules with multiple chapters and material releases and aligning it with the constraints of VLE data. This metric aggregates dimensions of student engagement—frequency, immediacy, and diversity of study sessions—into a singular value for the academic year.

For \textbf{RQ2}, we engineer predictors that reflect students’ weekly interactions with module resources, following methods used in previous studies \cite{nagrecha2017,chen2020,kloftetal2014}. We model the student engagement metric using regression models known for their performance with similar data, including penalised regression models, SVR, random forests, and GAMs. By comparing the performance of these models, we highlight that GAMs offer superior predictive accuracy and enhanced interpretability.

In addressing \textbf{RQ3}, we apply our methodology to a case study from University College London's (UCL) Department of Statistical Science. Using Moodle data from a first-year undergraduate module delivered in online, hybrid, and in-person modes over three years, we incorporate the delivery method as a predictor in the models. We identify the critical weeks and resources to predict student engagement by fitting, evaluating, and interpreting these models while controlling for the delivery method. This case study demonstrates the applicability and effectiveness of our modelling approach, providing insights into the dynamics of student engagement across multiple teaching environments.
 
\subsection{Paper Structure}
 
The remainder of this paper is organised as follows. In Section \ref{sec:SEM}, we detail the development of the student engagement metric, explaining adaptations made to existing metrics. Section \ref{sec:meth} outlines our modelling approach, including regression models and feature engineering. Section \ref{sec:caseresults} presents a case study from UCL's Department of Statistical Science, applying our methodology to a first-year undergraduate module delivered in online, hybrid, and in-person modes over three years. Section \ref{sec:discussion} discusses our findings within the broader context of student engagement research and suggests directions for future work. Finally, Section \ref{sec:conclusion} summarises our key findings and contributions.

\section{Student Engagement Metric}
\label{sec:SEM}
In this paper, we develop a student engagement metric based on the score proposed by \citeA{wong&chong2018}, which defines student engagement using immediacy, duration and frequency. Immediacy measured the time between resource release and student access, aligning with synchronous and timely engagement. Duration tracked the total study time, and frequency counted the number of student study sessions. The research concluded that the score effectively measured student engagement and identified at-risk students. Additionally, there was a direct correlation between the score and student grades, reinforcing that active engagement improves academic performance. \citeA{bakeretal2025} supported these findings, demonstrating a correlation between immediacy and student performance as a measure of student engagement. In contrast, \citeA{hoffmanetal2023} found that frequency was insufficient in isolation, thus indicating the necessity for a combination of factors.  

This score is a promising advance in defining student engagement through log data, yet it has limitations. Firstly, it is not universally possible to accurately record the duration of a student study session from VLE data. While we could use the time between logs as an estimate, we might not capture the time a student spends reviewing resources, such as viewing videos or reading lecture notes, and studying within an external environment.

In response to this limitation, we propose diversity as an alternative measure to duration. This factor counts the range of activities students interact with within the VLE, including accessing module resources, viewing and posting in the forum, and clicking on the URL of external resources. We argue this is a proxy for student engagement as it indicates an active participation in their learning process \cite{khanetal2017}, reflects intellectual curiosity and student agency \cite{reeve&tseng2011}.

A further limitation of the original scoring method was its assumption of a single material release, which students worked through independently. This approach overlooks the fact that resources are typically released weekly by chapter in higher education institutions. Thus, we need to extend the score to encapsulate this structure. To address this, we calculate a score for each chapter release and then aggregate these scores into a final student engagement metric using a weighted sum. The weighting is determined by each chapter's relative importance and learning objectives, ensuring a comprehensive and representative assessment of student engagement throughout the module.

Therefore, we quantify student engagement through this metric, $y^{(i)}$, for each student, $i$, which is calculated by  
\[y^{(i)} = \sum_{k} w_k IDF_k^{(i)} \ ,\]
where $w_k$ are the weights for chapter $k$, and $IDF_k^{(i)}$ is the engagement score for each chapter. This score is calculated by
\[IDF_k^{(i)} = I_{scaled,k}^{(i)} + D_{scaled,k}^{(i)} + F_{scaled,k}^{(i)}\ , \]
where $I_{scaled,k}^{(i)},\ D_{scaled,k}^{(i)},\  F_{scaled,k}^{(i)}$ are the scaled immediacy, diversity and frequency scores for chapter $k$, detailed in section \ref{partb}.

\subsection{Study sessions}
We derive the student engagement metric scores by analysing periods of active student engagement with course modules, known as \textit{study sessions}, as evidenced by the VLE log data. Identifying these study sessions is challenging as no clear markers within the VLE logs signal the beginning and end of engagement periods. \citeA{munk&drlik2011,valle&duffy2009,omaretal2007} suggest using a period of inactivity -- typically between 15 to 30 minutes -- to delineate the end of one session and the start of another, while \citeA{akpinaretal2020} advocates for adjusting the inactivity threshold based on the type of activity. Our empirical analysis found minimal differences between both approaches, leading us to adopt the simpler, fixed-threshold method for its straightforward applicability.

To determine this threshold, we examined the distribution of inactivity periods up to two hours, presuming that durations beyond this point signalled a new session. We choose the 95th percentile of these periods as the cutoff to effectively capture the bulk of engagement activity without conflating distinct study sessions. This method ensures that our engagement metric accurately represents the frequency of students' active learning.

Sessions are categorised by chapter, as indicated by accessed resources. Henceforth, \textit{chapter} $k$ refers to the studied chapter in a given session. When a student studied multiple chapters during the same session, we split the session into separate sessions for every chapter studied. For each chapter $k$ and student $i$, we collate information across all study sessions to derive three critical pieces of data: the number of study sessions denoted $L_k^{(i)} \in \mathbb{N}_0$; the earliest session date, denoted $\delta_k^{(i)} \in \mathbb{N}_0$; and activities engaged in, denoted by the binary vector $\mathbf{A}_k^{(i)} \in \{0, 1\}^{N_k}$. 

We determine the earliest session date by finding the minimum date across all sessions for chapter $k$, measured as the number of days from the start of term one. Formally, this is
\[
\delta_{k}^{(i)} = \min \{\delta_{k,l}^{(i)} \mid l = 1, \ldots, L_k^{(i)}\} \ ,
\]
where $L_k^{(i)}$ is the total number of sessions for student $i$ in chapter $k$ and $\delta_{k,l}^{(i)}\in \mathbb{N}_0$ is the date of each session $l$ in number of days.

The binary activity vector, \(\mathbf{A}_{k}^{(i)} \in \{0, 1\}^{N_k}\), indicates whether student \(i\) engaged in each of the \(N_k\) activities for chapter \(k\) at least once. This vector is constructed by aggregating the study session binary activity vectors, \(\mathbf{A}_{k,l}^{(i)}  \in \{0, 1\}^{N_k} \), across all \(L_k^{(i)}\) sessions, where a value of 1 indicates the corresponding activity was engaged in during session $l$, and 0 otherwise. We aggregate these vectors by taking the maximum value for each activity across all sessions, applying an element-wise maximum operation, which we express as
\[
\mathbf{A}_k^{(i)} = \max \{ \mathbf{A}_{k,l}^{(i)} \mid l = 1, \ldots, L_k^{(i)}\} \ .
\]

\subsection{Constructing the student engagement metric}
\label{partb}

We calculate immediacy and frequency scores for every student $i$ and chapter $k$ by adapting the definitions given by \citeA{wong&chong2018} while we developed the diversity score. For all three scores, a higher value indicates a higher level of student engagement.

Immediacy is the difference between the release date and the date of the first study session for each chapter, expressed as a negative value. Thus, an immediacy score of zero indicates that a student accessed the chapter resources on the day they were released. Conversely, a negative value signifies the number of days the student delayed before interacting with the resources after their release. The more negative the score, the greater the lag time between the release and the initial engagement. We determine the release date by identifying the first date any student accesses a resource from that chapter for that year. Therefore, immediacy is given by
\[I_k^{(i)} = \min_{i^\prime} \left( \delta_k^{(i^\prime)} \right) - \delta_k^{(i)}, \]
where $i^\prime$ are all students from the same academic year.

We define frequency as the number of study sessions per chapter for each student. Thus, the frequency score of student $i$ for chapter $k$ is defined as
\[ F_k^{(i)} = L_k^{(i)},\]
which is the total number of sessions recorded for student $i$ and chapter $k$. 

Diversity measures the variety of activities students engage in while studying each chapter. We quantify this metric by counting the distinct types of activities undertaken, as indicated by the non-zero elements in the binary vector $\mathbf{A}_k^{(i)}$. Formally, the diversity score for chapter $k$ and student $i$, denoted as $D_k^{(i)}$, is given by
\[ D_k^{(i)} = \sum_{j=1}^{N_k} \left( A_{k}^{(i)} \right)_j,\]
where $N_k$ is the number of distinct activities in chapter $k$. 

Min-max scaling is applied to all three scores, transforming the values to a 0-1 scale by
\[ I_{scaled,k}^{(i)} = \frac{I_k^{(i)}-\min I_k^{(train)}}{ \max I_k^{(train)}-\min I_k^{(train)}} \ , \]
where we calculate the minimum and maximum values for each chapter $k$ across students in the training set, similarly applied for $D_{scaled,k}^{(i)}$ and $F_{scaled,k}^{(i)}$. As the maximum and minimum values are determined in-sample, out-of-sample values can be outside the 0-1 range. However, this is not problematic as we will sum these values to construct the student engagement metric. This normalisation process allows for fair comparisons by adjusting scores to a consistent scale across each chapter while simplifying the integration of different engagement dimensions into a single metric. Furthermore, this approach aligns with existing research that conceptualises student engagement as a comparative measure \cite{kitturetal2022}, highlighting the importance of evaluating students' behaviour relative to their peers. 

We calculate the student engagement metric per student per chapter, $IDF_k^{(i)}$, by summing the scaled immediacy, frequency and diversity scores with equal weight. Finally, we derive the overall student engagement metric, $y^{(i)}$, as a weighted sum of $IDF_k^{(i)}$ across all chapters. 

\section{Methods}
\label{sec:meth}
This section presents our approach in four parts: defining nine regression models, transforming the student engagement metric to be the response variable for the models, describing how predictors are constructed from VLE data to capture students' weekly interaction with module resources, and specifying the criteria for evaluating model performance. Section \ref{sec:caseresults} that follows will apply this method to a case study to compare and interpret the models. 

\subsection{Models}
Among the regression models documented for their effectiveness in modelling student engagement and VLE data, we have chosen seven for their ability to manage multicollinearity among predictors alongside those that can account for non-linearity in the data. We also propose that a GAM is a suitable model for this data. We fine-tune each model's set of hyperparameters through 10-fold cross-validation during training. 

\subsubsection{Principal Component Regression (PCR)}

Principal components reduce the dimensionality of the predictors by capturing the maximum variance through linear combinations of the original continuous predictors. PCR then employs these principal components as continuous predictors in a linear regression model. Although the coefficients in the model can reveal the significance of the principal components, interpreting the model in terms of the original predictors can be challenging. 

\subsubsection{Penalised regression}

Penalised regression fits the standard linear regression model while adding a penalty term to the loss function to shrink the coefficients of less important predictors towards zero. This penalisation reduces overfitting and improves model generalisation. 

Ridge regression forces all coefficients to shrink uniformly towards zero. The relative magnitude of the coefficients provides insights into each predictor's significance. Conversely, lasso regression allows coefficients to shrink to precisely zero, not just close to it. This property inherently performs predictor selection, emphasising the most influential predictors in the model. Consequently, the interpretation of the lasso model hinges on identifying which predictors retain non-zero coefficients. Elastic net regression combines the strengths of ridge and lasso regression.

\subsubsection{Generalised Additive Model (GAM)}

GAMs maintain the framework of generalised linear models but fit non-linear smooth functions for each predictor. Initially, we will fit the model with all available predictors to assess their contribution to the model. We then conduct feature selection based on the significance of predictors across the training folds, retaining predictors found to be significant at the 0.1 level in over 50\% of the folds for the final reduced GAM, which we call rGAM. We give a detailed interpretation of this model in section \ref{interpretation}.

\subsubsection{Support Vector Regression (SVR)}

The kernel of SVR determines the transformation of the data. The linear kernel SVR behaves similarly to linear regression, meaning we can interpret the model through the weights which suggest each predictor's importance. The Radial Basis Function (RBF) kernel SVR can capture non-linear relationships by mapping the data into an infinite-dimensional space and fitting a linear regression function equivalent to a non-linear regression function in the original space. The transformation makes the model inherently complex, limiting direct interpretability. As such, the primary purpose of this model in the analysis will be performance benchmarking. If the RBF kernel SVR substantially outperforms the linear regression models, it may suggest potential violations of the linear relationship assumption between the predictors and the response variable.

\subsubsection{Random Forest}

Random Forest is an ensemble learning method that builds multiple decision trees during training and takes the mean of their outputs for prediction. Random forest is a non-parametric algorithm that does not make assumptions about the data. We can employ predictor importance scores to reveal the significance of each predictor in the model \cite{hastieetal2009}.

\subsection{Response variable}
We use the student engagement metric as the response variable in each model. The linear regression models presuppose homoscedasticity and normally distributed errors. To address potential deviations from these assumptions, we consider the application of a Box-Cox transformation to the student engagement metric. The Box-Cox transformation is a well-established technique for variance stabilisation and making distributions more symmetric \cite{boxcox}, described by the transformation rule
\[
y^{(i)}_{boxcox} (\lambda) = 
\begin{cases}
    \frac{(y^{(i)})^\lambda-1}{\lambda} & \text{if } \lambda \ne 0 \\
    \log(y^{(i)}) & \text{if } \lambda = 0,
\end{cases}
\]
where $\lambda$ is a hyperparameter to be optimised. 

\subsection{Model predictors}
We choose the models' predictors to reflect the weekly behaviour of the students. This approach is similar to the models developed for predicting student attrition as mentioned in \citeA{nagrecha2017,chen2020,kloftetal2014}. These predictors will highlight which specific weeks and resources are crucial in discerning high or low student engagement levels throughout the academic year. 

Suppose the module runs for \( T \) weeks and has a set of core resources, denoted as \( \mathcal{R} \). We define \( X^{(i)}_{r,t} \) to be the number of times student \( i \) accessed resource type \( r \) in week \( t \) of the module. This formulation leads to the set \( \{ \{ X_{r,t}^{(i)} \}_{r \in \mathcal{R}} \}_{t=1}^T \) of continuous predictors for each student \( i \).

\citeA{akpinaretal2020} also suggests that counting the number of clicks within the VLE each week might also be a strong predictor of student behaviour. Thus we can include $\{ X_{C,t}^{(i)} \}_{t=1}^T$ in the predictors, where $C$ denotes clicks, which counts the total number of logs within the VLE data for student $i$ in week $t$.

Standardising continuous predictors is necessary for models sensitive to scale, including penalty regression and SVR. We centre each predictor around zero with a standard deviation of one. The training set determines the parameters of the standardisation, such that
\[\Tilde{X}_{r,t}^{(i)} = \frac{X_{r,t}^{(i)} - \bar{X}_{r,t}}{ \hat{\sigma}_{X_{r,t}}}\]
where $\bar{X}_{r,t}$  is the mean and $\hat{\sigma}_{X_{r,t}}$ is the standard deviation of $X_{r,t}^{(i)}$ across all students in the training set for week $t$. The scaled predictors are used in every model to ensure a direct comparison between model performances. Thus, 
\[ \{ \{ \Tilde{X}_{r,t}^{(i)} \}_{r \in \mathcal{R}},\ \Tilde{X}_{C,t}^{(i)} \}_{t=1}^T\]
is the set of standardised continuous predictors for student $i$ used in every model.

Alongside these weekly behaviours, the module delivery method experienced by the student, $G^{(i)}$, is also included, such that
\[G^{(i)} = 
\begin{cases}
  1 & \text{if online} \\
  2 & \text{if hydrid} \\
  3 & \text{if in-person.} \\
\end{cases} \]

To ensure our approach adapts to different contexts, we can feasibly omit this variable or integrate additional interesting variables, including changes in the instructional team, curriculum adjustments, or other significant factors, to understand their impact on student engagement.
\label{pred}

\subsection{Model evaluation}
We train our models using nested 10-fold cross-validation. The outer stage reduces the risk of overfitting and allows us to evaluate our models using the out-of-sample root mean square error (RMSE) and $R^2$, while the inner stage tunes the hyperparameters through grid search. Once we find the optimal hyperparameters and evaluate the models across all ten folds, we retrain the models on the full dataset to leverage the entire dataset to interpret the models.

We can further understand each model's performance through segment analysis. To achieve this, we categorise the samples into quintiles according to their student engagement metric and calculate each model's RMSE and mean error for each group. This analysis gives insight into the model's accuracy for different levels of student engagement. In particular, we will be interested in which models most accurately identify students displaying low engagement levels, as these students might require additional support.

\section{Case Study and Results}
\label{sec:caseresults}
In the following section, we apply our methodological framework to a specific case study focusing on the module STAT0004: Introduction to Practical Statistics, offered by the Department of Statistical Science at UCL. The STAT0004 module is compulsory for first-year statistics undergraduates and imparts statistical coding skills in the R programming language \cite{R}. The module runs across two terms, from university week 5 to week 35, with a summative quiz in week 16 and a group coursework due in week 35. The winter break (weeks 17-19) separates terms one and two. Students work independently in term one (weeks 5-16) and transition to assigned groups in term two (weeks 20-30).

The module uses a flipped classroom structure through three primary resources complemented by practical coding sessions. These resources, accessed on the Moodle site, include lecture notes for each chapter, videos supporting the lecture note content, and formative quizzes. Additionally, students access external resources, a forum, and Zoom meetings for remote coding labs through the Moodle platform. 

We chose this module due to its mixed mode of delivery over three years. In the academic year 2020-2021, the COVID-19 pandemic led to the adoption of remote learning, conducting all labs and interactions via Zoom or the Moodle platform. In the academic year 2021-2022, a hybrid delivery model was implemented, with some labs conducted in person and others online. All labs were held in person during the academic year 2022-2023, marking a return to the standard module delivery after the disruptions caused by the pandemic. Moodle has been the primary channel for learning resources and course communication throughout all three years. We have leveraged this unique set of circumstances to understand the impact of delivery methods on student engagement. However, we must note that there was a different cohort of students each year and differing contexts in the broader world regarding the pandemic. Therefore, other unobserved variables likely confound the impact of the delivery method.

\subsection{Case Study: The data}
We extracted data from the UCL Moodle platform for the STAT0004 module for three academic years. The dataset has six columns that provide details such as the click time and the type of action the user takes. Table \ref{table_datcolumns} describes each column.

\begin{table}[hbt]
\centering
\renewcommand{\arraystretch}{1.3}
\caption[Moodle dataset]{\label{table_datcolumns}
Summary of the Moodle activity log data, highlighting the key columns and the information they contain.}\vspace*{1ex}
\begin{tabular}{l p{9cm}}
\hline
Column name & Column description \\[1ex]
\hline
\texttt{Time} & The data and time (to the nearest minute) \\
\texttt{User} & User ID number for anonymity  \\
\texttt{Event.context} & Title of the link or resource accessed \\
\texttt{Component} & The category of activity or resource accessed \\
\texttt{Event.name} & The purpose of the click \\
\texttt{Description} & Details of the click, including user ID and resource ID \\[1ex]
\hline 
\end{tabular}
\end{table}

Only students enrolled on the course for the first time are included in the analysis, giving $N=709$ unique users. Furthermore, we restrict the data to the period between university weeks 5 and 35 for each academic year. We identify lecture notes, videos, and quiz accesses by the text in the \texttt{Event.context} and \texttt{Event.name} columns, extracted using the \texttt{stringr} R package \cite{stringr}.

\subsection{Case study: student engagement metric}

The 95th percentile of the lag times between logs is approximately 45 minutes (to the nearest 5 minutes). Therefore, we determined that a cut-off interval of 45 minutes for a study session is optimal; any lags beyond this duration are considered the start of a new session. Although this duration exceeds recommendations in some literature, we can attribute it to the module's nature, which demands that students practice their programming skills in an external R programming environment, which might cause longer lags in the logs.

We determine the number of activities for the diversity metric by the unique combinations of entries from the \texttt{Event.context}, \texttt{Component} and \texttt{Event.name} columns. Chapters 1--10, studied in term one, are assigned a weight of 0.6, while chapters 11--20, term two, are assigned 0.4 to reflect that students are working in groups. 

The Box-Cox transformed metrics have values between 0.5 and 30.2 and a mean of 14.9. There are slight discrepancies in the distribution of the engagement metric across the different delivery methods. Specifically, the mean of the transformed metric is 16.2 for online, 14.6 for hybrid, and 13.2 for in-person, suggesting a consistent decrease in engagement year over year.

\subsection{Case study: model predictors}

We count four behaviours for each $t=5,\ldots,35$ weeks for each student $i$, reflecting the interaction with the three core resources for the module alongside the total number of clicks within Moodle. Thus, we have the total number of
\begin{itemize}[noitemsep]
\item clicks, $X_{C,t}^{(i)}$
\item lecture note accesses, $X_{L,t}^{(i)}$
\item video views, $X_{V,t}^{(i)}$
\item quiz submissions, $X_{Q,t}^{(i)}$ 
\end{itemize}
where $X_{Q,5}^{(i)}=0$ for all $i$, as the first quiz was released in week 6.

Across the three delivery methods, one lecture note and quiz were consistently available per chapter for each of the 20 chapters. However, the availability of videos differed: the fully remote year featured 45 videos, whereas the hybrid and in-person years provided only 28. Despite this variation, videos served a consistent supplementary role across all delivery methods, offering detailed explanations that complemented the lecture notes. Therefore, for a fair comparison, the weekly video view predictors are transformed into a proportion of the total number of available videos per year, giving proportional video views
\[X_{V,t}^{\prime(i)} =
\begin{cases}
  \frac{X_{V,t}^{(i)}}{45} & \text{for $i$ such that $G^{(i)} = 1$} \\
  \frac{X_{V,t}^{(i)}}{28} & \text{for $i$ such that $G^{(i)} = 2,3$.} \\
\end{cases} \] 
After removing the sparse predictors (less than 1\% non-zero entries), there are $p=118$ predictors, including the discrete delivery method variable. 

\subsection{Case study: exploratory data analysis}

\begin{figure}[ht]\centering
\includegraphics[width=\linewidth]{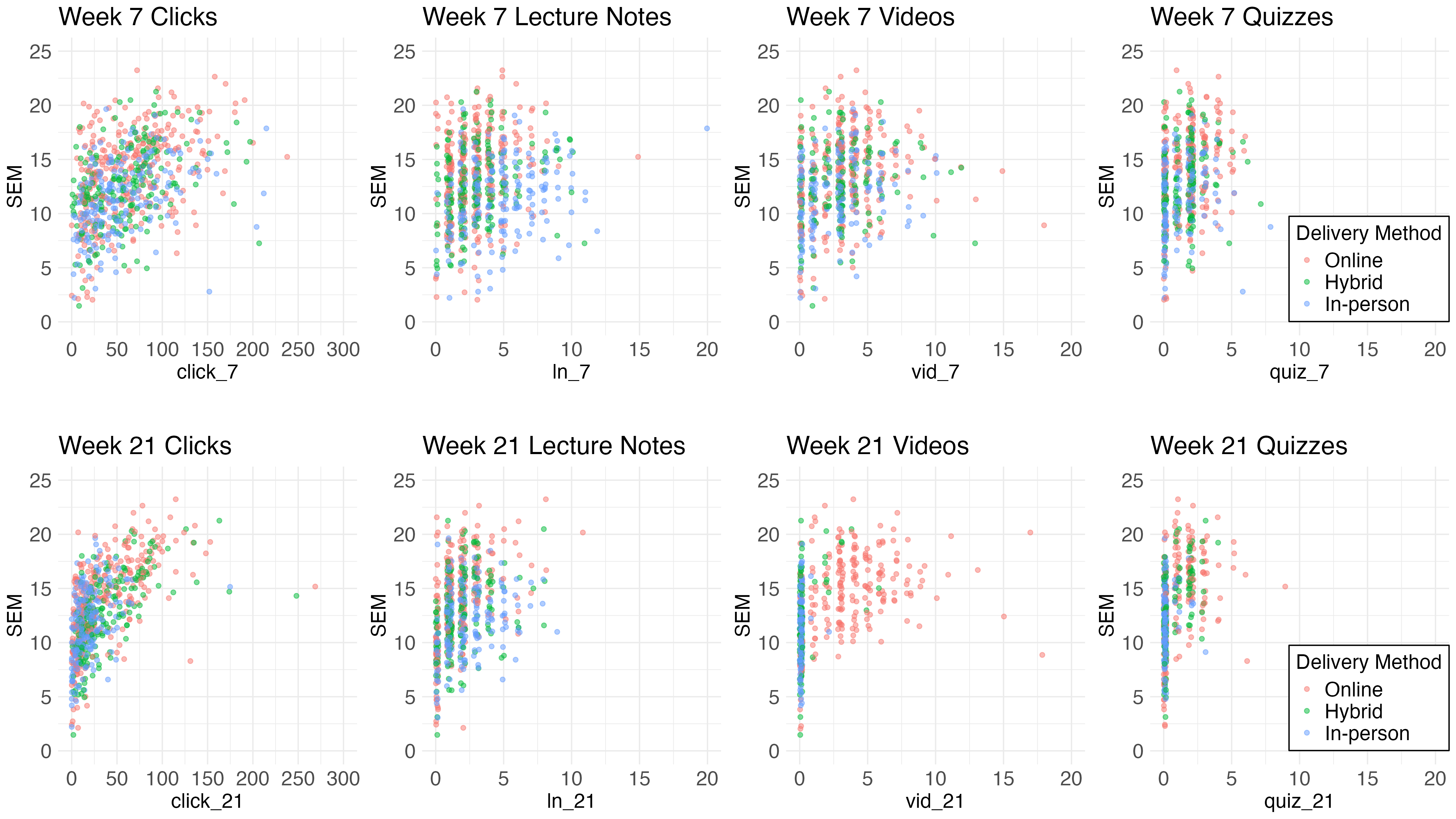}
\caption{Scatter plots illustrating the relationship between the student engagement metric and the four predictors — clicks, lecture note accesses, video views, and quiz submissions — for weeks 7 (term 1) and 21 (term 2). Each point represents a sample colour-coded by the delivery method (online, hybrid, in-person).}
\label{fig_trends}
\end{figure}

Visualising the weekly continuous predictors offers insights into their relationship with student engagement and the overall data structure. Figure \ref{fig_trends} illustrates scatter plots for weekly clicks and resource interactions for weeks 7 and 21, mapped against the student engagement metric, colour-coded by delivery method.

A positive correlation emerges across all plots, indicating higher engagement levels with increased VLE clicks and resource interactions. Despite this general trend, the influence of the delivery method on these correlations varies. For example, week 21's video views predominantly feature online students (marked in red), suggesting a delivery method-dependent engagement pattern, possibly due to the change in video availability. Although some plots indicate independence between the predictor and the delivery method, such as the evenly distributed clicks in week 7, other instances suggest a dependency. Notably, in-person students (blue data points) in week 21 had fewer clicks, whereas in week 7, they accessed the lecture notes more often.

Comparing weeks 7 and 21 reveals a decline in VLE usage and resource access. This pattern is apparent across the complete set of predictors, which displays a considerable decrease in activity from term one to term two. This decline in activity is likely, at least in part, due to students working in groups in term two but might also signify a decline in engagement post-winter break. 

Quizzes emerge as the least engaged resource, likely due to the nature of quizzes requiring only a single attempt per chapter. This observation may also reflect student preference amongst resources, viewing quizzes as optional or of lesser importance, suggesting attitudes towards formative assessments.

\subsection{Case Study: Model evaluation}
\begin{figure}[ht]\centering
\includegraphics[width=4in]{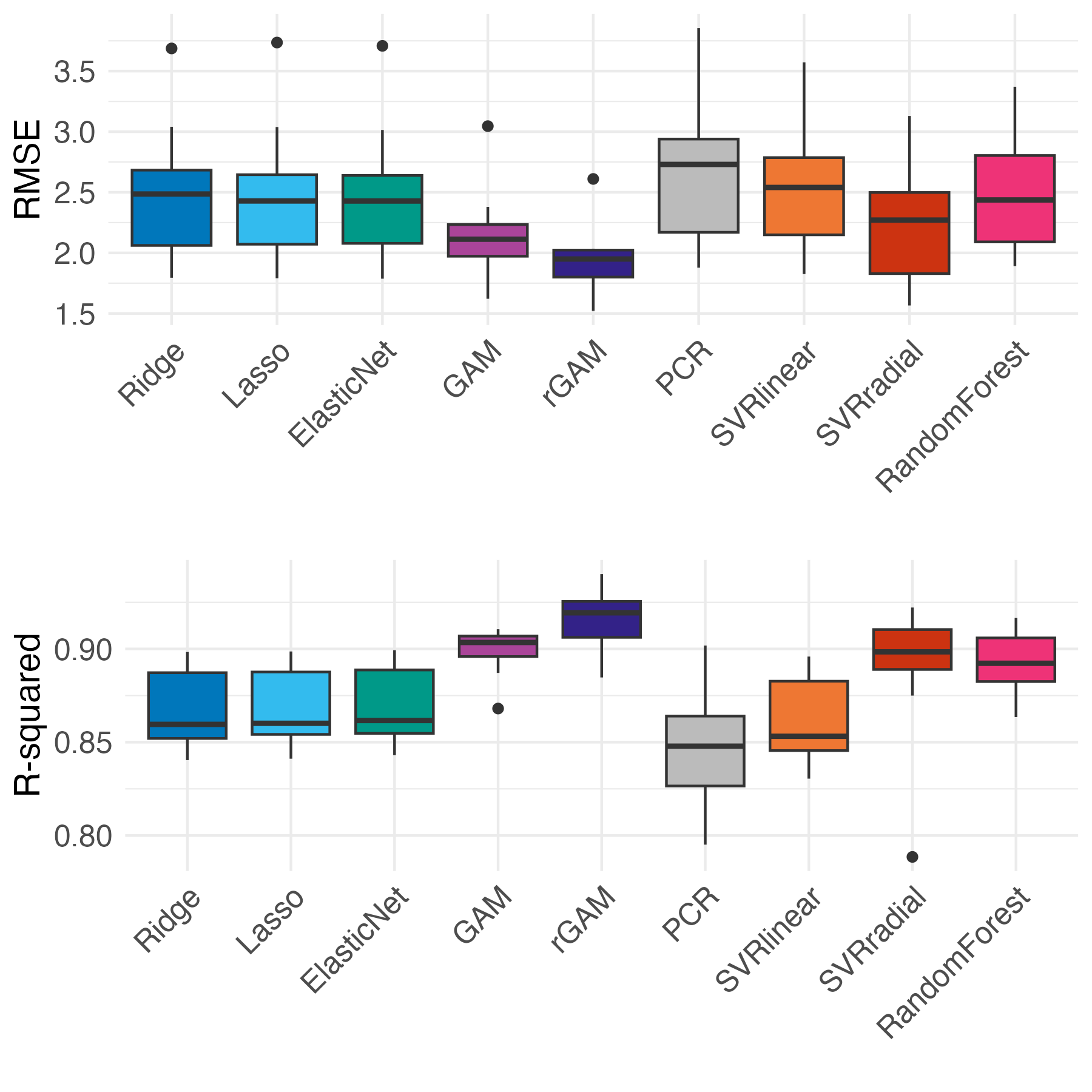}
\caption{Boxplots comparing the out-of-sample RMSE and R-squared values across ten folds for the nine regression models.}
\label{fig_performance}
\end{figure}

Figure \ref{fig_performance} shows the boxplots of model performance metrics for the out-of-sample data across ten folds. Penalty regression models (Ridge, Lasso, ElasticNet) and linear SVR display similar RMSE and $R^2$ values, indicating comparable effectiveness in predicting student engagement. The PCR model, however, shows slightly higher RMSE and lower $R^2$ values, suggesting a marginal reduction in performance. 

The models capturing non-linear relationships performed better. The full GAM, SVR with RBF kernel, and Random Forest models have comparable $R^2$ means and spread, all showing an improvement compared to the linear models. However, the reduced GAM (rGAM) surpasses them all, with the lowest mean RMSE (1.93) and the highest mean $R^2$ value (0.92), highlighting the importance of feature selection in reducing noise and improving model accuracy in predicting student engagement.

\begin{figure}[ht]\centering
    \includegraphics[width=4in]{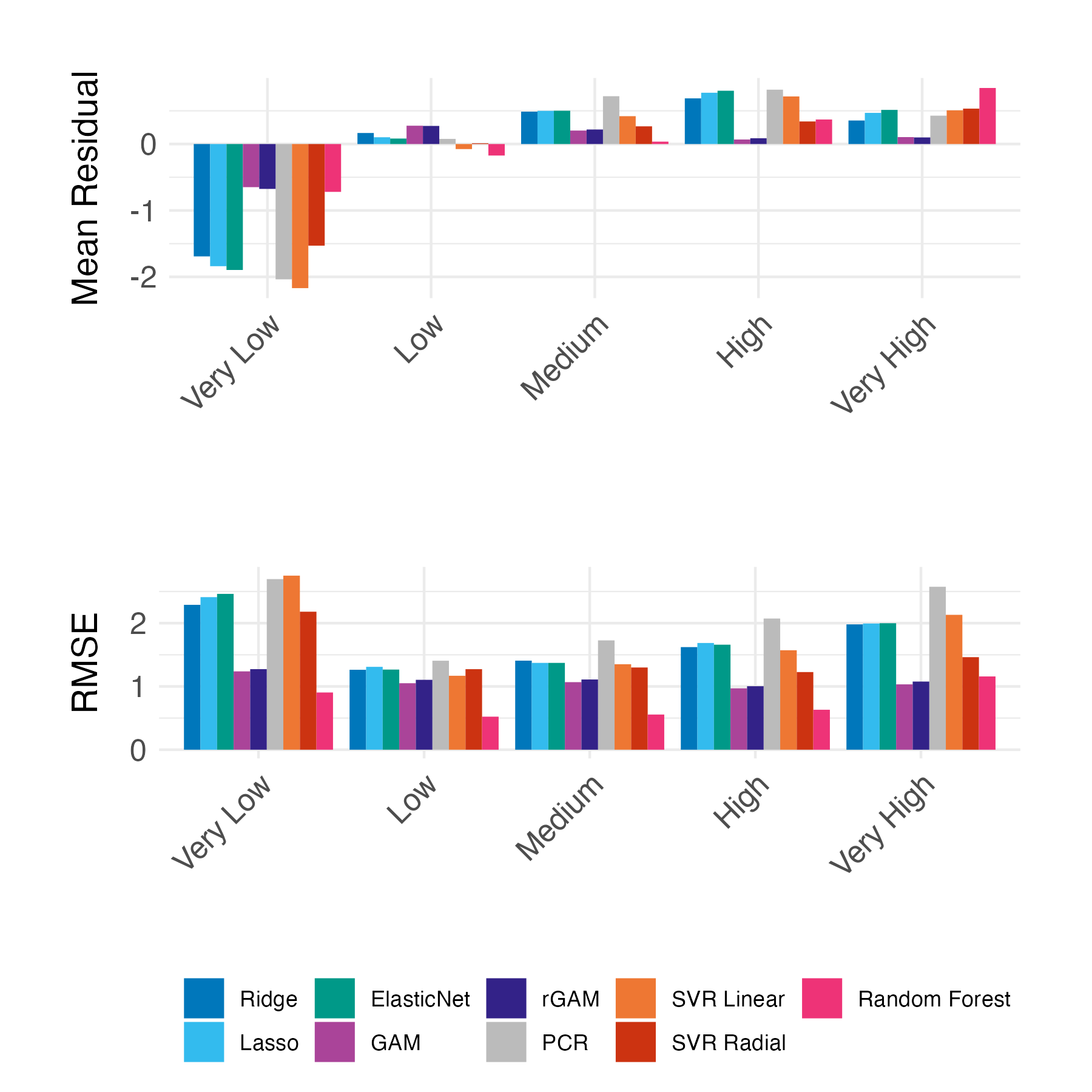}
    \caption[Large Residuals bar plots]{Bar plots showing each final model's mean residual and RMSE, segmented by student engagement levels. We assign student engagement levels according to the quintiles of the student engagement metric (very low - very high).}
    \label{figure_segmentplots}
\end{figure}

Figure \ref{figure_segmentplots} displays the final models' performance segmented by student engagement levels, with mean residuals indicating the average deviation of predictions from observed values. The figure shows that models tend to overestimate engagement for students at lower engagement levels and underestimate it for those at higher levels. This pattern results in predictions clustering towards the mean, with RMSE values indicating greater prediction errors at the extremes of student engagement. Overestimating the lowest engagement group is more of a concern than underestimating the highest group, as this could hinder the accurate identification of at-risk students, posing a challenge to targeted intervention efforts. 

The results reinforce the notion that the two GAM models exhibit strong performance in predictive accuracy. Alongside the GAM models, the Random Forest model stands out for its consistent accuracy across the engagement spectrum, demonstrated by the lowest RMSE values for four segments of students (very low - high engagement levels). Critically, these three models achieve the mean residual closest to zero for the very low engagement group, underscoring their precision in identifying at-risk students.

Given these insights, our analysis will focus on the rGAM and Random Forest models. These models perform well across all engagement levels and offer interpretive value.

\subsection{Case Study: Interpretation of the models}
Our analysis aims to identify the critical weekly behaviour predictors of student engagement and examine how delivery methods impact this engagement. By comparing findings from the rGAM and Random Forest models, we aim to validate the consistency of our results. In the Random Forest model, we use permutation to determine the importance of each predictor and evaluate its impact on model accuracy. This approach allows a direct comparison with predictor significance in the rGAM model. We then interpret the influence of each predictor in the rGAM model on the response variable by plotting its smooth functions to illustrate the nature of the non-linear relationships. 

\subsubsection{Continuous predictors}

\begin{figure*}[ht]\centering
    \includegraphics[width=\textwidth]{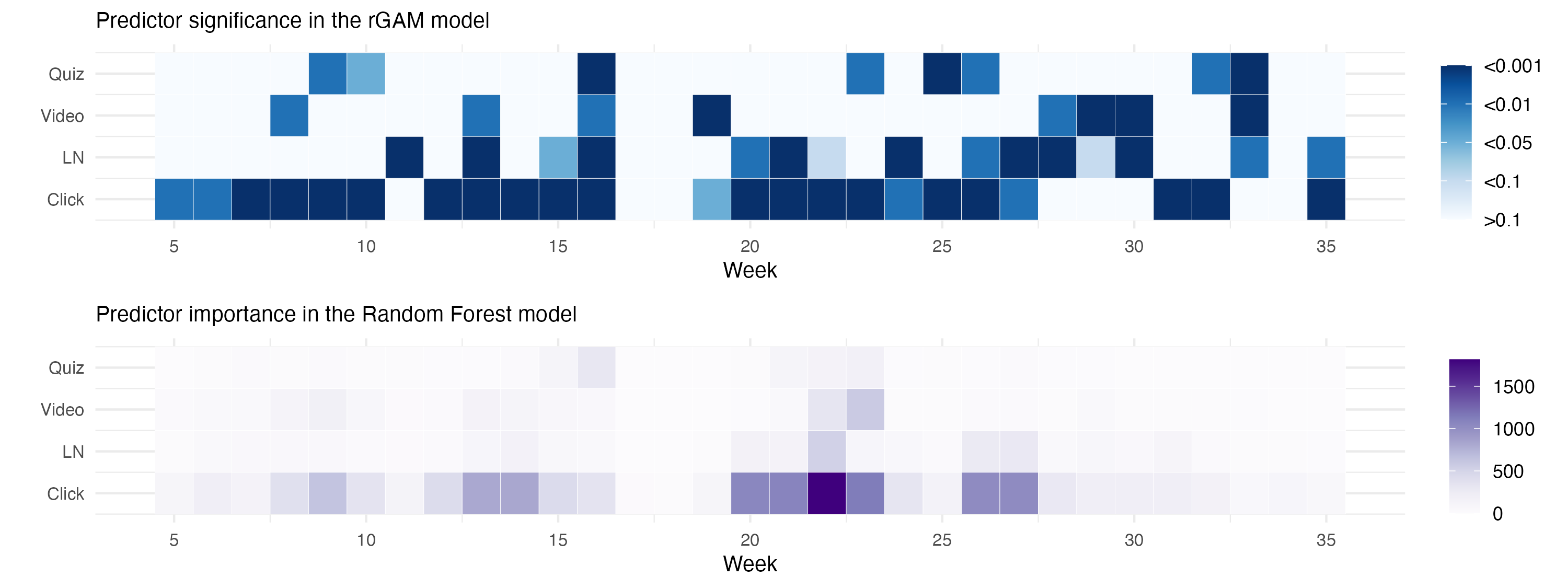} 
    \caption[Coefficient heatmaps]{Heatmaps illustrate the significance of predictors in the rGAM and variable importance in the Random Forest model. In the rGAM heatmap, colours represent the significance level of each predictor according to its p-value; darker shades indicate higher significance. White indicates predictors that were either insignificant or not included in the model. For the Random Forest heatmap, the right-side legend denotes the importance of each predictor, with comparative magnitudes offering more insight into their relative importance than absolute values. This visualisation aids in understanding which predictors are most influential across both modelling approaches.}
    \label{figure_coefplots}
\end{figure*}

Figure \ref{figure_coefplots} presents a heatmap comparing the significance of continuous predictors in the rGAM model with their importance scores in the Random Forest model, organised by resource type and week. In the rGAM analysis, predictors are denoted significant by their p-values: darker shades correspond to lower p-values and higher significance. Specifically, 54 predictors demonstrate a $\text{p-value} < 0.1$, all of which we retain in the reduced model, with 52 showing significance at the 0.05 level. Contrastingly, in the Random Forest model, predictor importance is quantified between 0 and 1815, with the darkest shade representing the maximum score. Six predictors exhibit an importance score exceeding 1000, and only 11 surpass the 500 mark. This visualisation highlights key predictors and facilitates a direct comparison of predictor relevance across the two models.

\begin{figure}[ht]\centering
    \includegraphics[width=.75\textwidth]{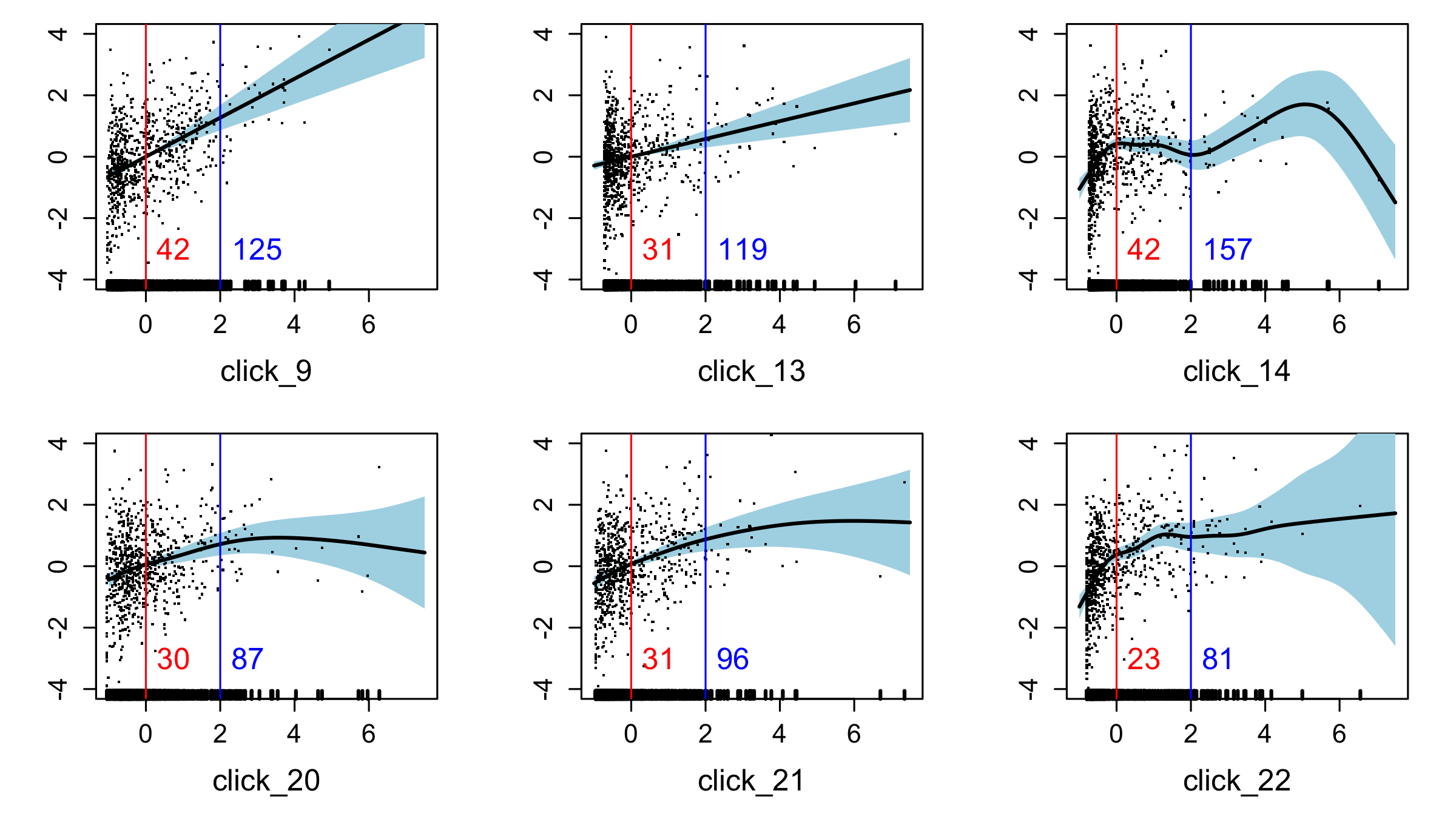} 
    \caption[Click smooth plots]{Smooth functions from the rGAM model for click predictors across weeks 9, 13, and 14 in term one, and 20, 21, and 22 in term two. Each curve represents the modelled relationship between the number of clicks and student engagement for the respective week, with confidence intervals shaded in light blue. The mean value before standardisation is labelled in red, while the value at two standard deviations above the mean is labelled in blue. A rug plot on the x-axis shows the distribution of data points.}
    \label{figure_clicksmoothplot}
\end{figure}

Figure \ref{figure_clicksmoothplot} illustrates the smooth function plots for select predictors from the rGAM model. These plots reveal the predictors' relationship with the response variable by describing their estimated effects with smooth curves and confidence intervals. It's essential to remember that the model's predictors were standardised to a mean of zero — marked by the red line — and a standard deviation of one as part of the model fitting process. Notably, the confidence intervals widen for values beyond two standard deviations from the mean — highlighted by the blue line - which indicates a reduction in prediction reliability at these more extreme predictor values.

Clicks are a consistently important predictor of student engagement throughout the first and second terms in the rGAM and Random Forest models. In the first term, the Random Forest shows the importance of clicks peaks notably in week 9 with a score above 600 and further in weeks 13 and 14, surpassing 800. The agreement on the high significance and importance of clicks during these weeks may signal an opportunity for targeted early intervention, with students identified through the number of clicks in the VLE.

In the second term, the importance scores for weeks 20, 21, and 23 each exceed 1000 and week 22 registers the highest score of 1815. This sustained relevance of clicks following the winter break suggests that how students re-engage with the course through the VLE is an essential indicator of their overall student engagement. Furthermore, during this period, students are working in groups. Therefore, it might also imply that individual engagement with the VLE remains critical, even as students collaborate. 

Examination of the rGAM model's smooth function plots for click predictors reveals two distinct patterns within the range of up to two standard deviations, where the model's predictions are most reliable. One pattern consistently increases, suggesting that more clicks are always associated with higher overall student engagement. This trend is observed in Figure \ref{figure_clicksmoothplot} for clicks in weeks 9, 13, 20, and 21. In contrast, the second pattern shows an increase in engagement up to the mean number of clicks, which then reaches a plateau, indicating that additional clicks do not correspond to increased engagement beyond a certain threshold. This latter pattern is particularly noticeable in week 14 and, to some extent, week 22. While this means that these weeks might not differentiate as well among students with higher numbers of clicks, all of these weeks have the potential to highlight less engaged students.

\begin{figure}[ht]\centering
    \includegraphics[width=\textwidth]{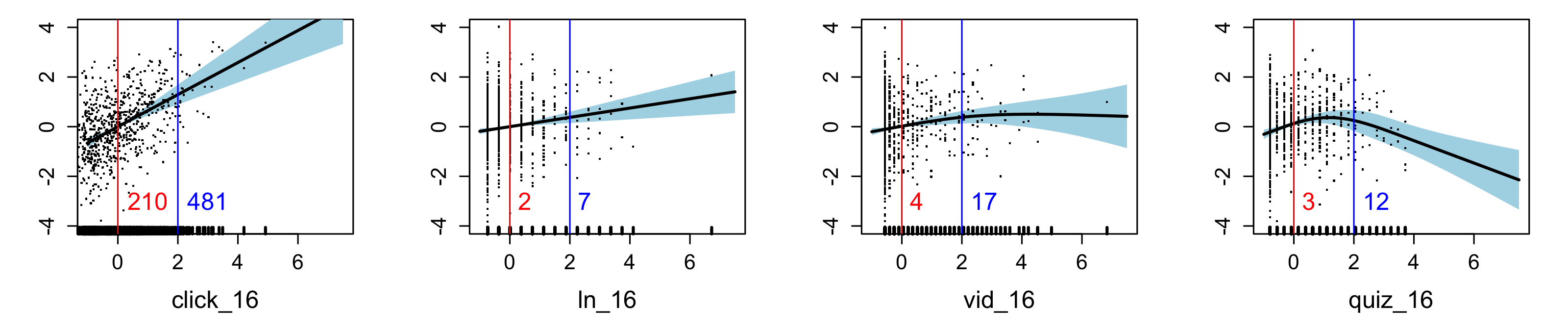} 
    \caption[Week 16 smooth plots]{Smooth functions from the rGAM model for all four predictors - clicks, lecture note (ln) accesses, video (vid) views and quiz submissions - for week 16.  Each curve represents the modelled relationship between the number of clicks or resource interactions and student engagement, with confidence intervals shaded in light blue. The mean value before standardisation is labelled in red, while the value at two standard deviations above the mean is labelled in blue. A rug plot on the x-axis shows the distribution of data points.}
    \label{figure_week16smoothplot}
\end{figure}

Week 16 stands out, as the rGAM model registers all four resources as significant, while the Random Forest analysis reveals the highest importance score for quizzes over both terms. Notably, this week coincides with the first summative assessment. Figure \ref{figure_week16smoothplot} shows the smooth plots for all four resources. For clicks, lecture notes (ln), and video (vid) views, the plots exhibit a consistent increase in the relationship between these interactions and overall student engagement, extending up to two standard deviations from the mean. Conversely, the positive correlation for quizzes extends only to one standard deviation; beyond this point, the correlation with quizzes diminishes. This pattern suggests an optimum level of engagement through quizzes that, when exceeded, may no longer contribute to or even detract from overall engagement.

The patterns underscore that diversified interaction with module resources, particularly in assessment preparation, indicates high student engagement overall. However, beyond a certain threshold of quiz submissions (approximately eight), excessive reliance on formative assessments close to summative assessment times could indicate catch-up behaviour, potentially signalling less consistent engagement in the preceding weeks. Overall, this highlights a change in student behaviour close to an assessment and the importance of this period regarding student engagement.


In the rGAM analysis, quizzes and video views are significant in only eight weeks throughout the academic year. Similarly, in the Random Forest model, quizzes are deemed the least impactful, highlighted by importance scores below 170, except week 16, which has a score of approximately 300. Video views only surpass an importance of 200 during weeks 22 and 23. This pattern of limited significance contrasts with much of the literature, which typically highlights quizzes \cite{akpinaretal2020,chen2020} and video views \cite{nagrecha2017} as critical indicators of student engagement.

The apparent reduced significance of quizzes and video views in our analysis may initially seem to underplay their importance in student engagement. However, this observation likely stems from their high collinearity with clicks — since accessing these resources requires clicking — leading the models to favour click metrics as more comprehensive predictors. Thus, the influence of quizzes and videos on engagement is implicitly included within click activity, suggesting their roles are not diminished but rather integrated into the broader metric of clicks prioritised by the models.

\subsubsection{Delivery method}

The discrete delivery method predictor analysis reveals a disagreement between Random Forest and rGAM. In the Random Forest, the delivery method predictor has a low importance score of approximately 8, indicating that removing this predictor would have minimal impact on the model's accuracy. In contrast, the rGAM approach ascertains the delivery method predictors as statistically significant at the 0.05 level. The effect estimates indicate that, with other predictors held constant, the model predicts a 0.3 and 0.6 unit increase in student engagement for hybrid and in-person methods, respectively, compared to online. However, the relatively small magnitude of these effects (0.3 and 0.6) suggests that while statistically significant, the practical significance of comparing delivery methods might be modest.  

We will discuss these results in section \ref{sec:RQ3}.
\label{interpretation}

\section{Discussion}
\label{sec:discussion}
This study aims to address three main research questions: 
\begin{description}
    \item[RQ1] How can student engagement be quantified using VLE data?
    \item[RQ2] How can student engagement be modelled using weekly interactions with resources as predictors?
    \item[RQ3] How can we interpret which weeks and resources are most important in predicting student engagement while accounting for the module’s delivery method?
\end{description}
We will discuss each of these in order in the following sections. 

\subsection{RQ1: Quantifying student engagement}
The student engagement metric introduced in this study encompasses multiple dimensions of engagement, reflecting a holistic approach supported by existing literature. We designed the metric to universally apply across different learning contexts and disciplines while accommodating sequential chapter releases. The metric summarises student engagement across an entire module and provides a comprehensive overview of student behaviour.

Nonetheless, the metric's reliance on VLE data introduces limitations, notably its potential not to capture engagements outside the VLE environment. This discrepancy underlines the inherent differences in capturing student behaviour between platforms like Moodle and MOOCs, which often offer more integrated learning environments. Thus, the metric may inherently reflect fully online learning contexts more than hybrid or in-person formats. One possible workaround for future research would be to incorporate more data from different sources, such as attendance data \cite{akpinaretal2020}, to achieve a more complete picture of student engagement.  

The metric's effectiveness during group work also presents challenges. Specifically, it may be difficult to ascertain whether it measures individual engagement accurately or inadvertently captures other group dynamics, such as instances where a single student's resource access might represent the activity of an entire group. This issue complicates the metric's utility in distinguishing genuine engagement from other types of participation, potentially conflating distinct behaviours. Consequently, while the student engagement metric serves as a valuable tool for summarising engagement, its capacity to encompass the entirety of student engagement, particularly beyond the VLE and within group activities, requires further evaluation.

\subsection{RQ2: Modelling student engagement}
According to the case study, the Random Forest model was found to be one of the most accurate models for the data, which is consistent with the literature \cite{akpinaretal2020,gorgunetal2022,bulathwela2020,sivaneasharajah2020,zhangetal2021}. The study also highlighted the potential of GAMs for modelling this type of data. The superior performance of these models underscores the complexity of the data, which requires non-linear models to capture the relationships between the weekly predictors and student engagement. It also suggests that there may be noise in the predictors caused by uninformative variables. Therefore, feature selection (performed manually for the rGAM and inherently in the Random Forest) is a vital processing step. 

The rGAM model offered a detailed interpretation of how resource interactions impact student engagement through smooth functions. The flexibility of these functions enables the identification of specific frequency ranges of interactions and their differential effects on engagement. With this approach, we can discover optimal activity levels beyond which engagement may plateau or even diminish.

While GAMs are an exciting prospect for future EDM research, they are not without their drawbacks. GAMs can be computationally intensive and time-consuming, especially when dealing with large datasets or many predictors. Consequently, the extended computation time renders traditional feature selection methods, such as forward or backward selection, impractical for use with GAMs. Given that feature selection not only enhances model performance but also aids in interpretability, it's crucial for future research to explore efficient yet effective feature selection strategies.

\subsection{RQ3: Predictors of student engagement}
\label{sec:RQ3}
The Random Forest and rGAM agree on several key predictors of student engagement. Firstly, both models identify early weeks in term one as critical for predicting engagement levels. This consensus may highlight the significance of establishing positive routines and student behaviours early in the module. Given the module's emphasis on independent study, these findings underscore the importance of early engagement as a precursor to successful module completion. Consequently, these critical weeks are interpretable as checkpoints for the lead lecturer to identify at-risk students and intervene early without waiting for more extended periods of disengagement.

Furthermore, the smooth functions revealed that the predicted level of engagement increased for most predictors until at least the mean. This insight reinforces the utility of comparing students against their peers \cite{kitturetal2022} as a practical benchmark for identifying at-risk students. An avenue for future research could explore how timely interventions during these pivotal early weeks could influence students' trajectories within the module.

Both models also pinpoint the week of the first summative assessment at the end of term one as a pivotal period, aligned with the findings of \citeA{akpinaretal2020}, who determined student behaviour adapted according to the assessment type students were preparing for. This finding suggests a trend of \textit{just in time} preparation for the summative assessment, in contrast to the extended preparatory phase that we might expect. Future research could leverage the grades of the first assessment \cite{golchehreh2023} alongside engagement patterns observed in the run-up to the assessment to identify students requiring additional support in term two.

Finally, both models also identified that the weeks at the beginning of term two, once students have returned from the winter break, are also highly important. This observation is intriguing as the construction of the student engagement metric assigned a smaller weight to chapters in term two. There are several possible explanations for this result. First, the student engagement metric encapsulates behaviours across all chapters, half of which are in term two. Therefore, to achieve a high value for the student engagement metric, students must interact with materials from all chapters on Moodle. The exploratory data analysis revealed that students significantly reduced resource interactions during term two. Therefore, these weeks might be more influential in discerning the different engagement levels due to the high variation in student behaviour.

Alternatively, the interpretation of the model might suggest that maintaining a certain level of independent study during group work is required to achieve high levels of student engagement throughout the year. It might also indicate that students who took active roles within the groups were more highly engaged than the other members.

It is crucial, however, to acknowledge the limitations in extrapolating these findings to different settings. The unique course structure, particularly the flipped classroom model and the computing discipline focus, might influence the engagement patterns observed. It would be necessary to integrate and analyse data from a broader array of module styles to ascertain the transferability of these results to other educational contexts, such as traditional lecture-based courses or non-computing modules. This expanded dataset would enable a more comprehensive understanding of how much these observed engagement trends are a product of the module design versus general principles of student engagement.

The Random Forest and rGAM models also occasionally diverged in assessing important predictors. In particular, they seemed to disagree on the significance of post-week 30 activities, which occur after the official end of the term while students complete the group coursework. The rGAM model highlighted these weeks as significant, while the Random Forest found a low importance for all predictors after week 27. On inspection, we found that the data for these weeks was relatively sparse, with a high percentage of samples equalling zero, indicating minimal student interaction with the VLE and resources during this period. Two possible explanations arise for the rGAM model's emphasis on post-week 30 activities despite the sparsity of data. First, the rGAM might genuinely detect significant relationships within the non-zero samples, linking them to distinct levels of student engagement during these weeks, thus using these insights to differentiate effectively. Alternatively, the model could overfit the limited data available for these periods, interpreting random fluctuations as meaningful patterns. In the data preprocessing stage, we set a cutoff of 1\% sparsity for including predictors in the model. Future research could optimise this sparsity threshold and consider the effect of narrowing the focus to term time with higher levels of recorded behaviours.

The findings on the effects of the delivery method on student engagement prove inconclusive, stemming from several methodological and contextual challenges encountered during the analysis. Firstly, the exploratory data analysis revealed the continuous predictors were not independent of the delivery method. We observed this in week 21, for example, where video views were considerably higher for students enrolled in online delivery modes, possibly due to the difference in availability of video resources across different delivery methods. This observation suggests the discrete predictor does not exclusively capture the relationship between delivery method and student engagement. 

Confounding variables, such as changes in the student cohort and the global context due to the COVID-19 pandemic, further complicate this shortfall. Furthermore, given that we collected our data through an online platform, the data might be inherently biased towards students enrolled when learning was fully online, as more interactions took place within the VLE. Considering these limitations, it is difficult to quantify the difference the delivery method has on overall student engagement. 

Despite these challenges, our study has highlighted observable changes in student behaviour when interacting with resources through the VLE year-on-year, suggesting another exciting avenue for future research. Integrating more data from various modules would enhance our understanding of these behavioural changes and help determine whether they result from the module's delivery method. This approach could then guide the development of more effective educational strategies tailored to specific contexts and delivery formats.

\section{Conclusion}
\label{sec:conclusion}
This paper presents a comprehensive measure of student engagement for a module, building upon the metric by \citeA{wong&chong2018} to accommodate VLE data and sequential material releases. We have shown that weekly VLE interactions can predict overall engagement levels by employing this student engagement metric as the response variable in nine regression models. Our results validate the effectiveness of the Random Forest model in modelling activity log data, as supported by the literature, and illustrate the superior performance and interpretive advantages of GAMs.

Through interpreting the models, we found that the total number of clicks within the VLE is generally a sufficient predictor of engagement for most weeks without the need to account for specific resource interactions. Early weeks within the first half of the first term were significant, indicating the potential for the early identification of students requiring additional support. Additionally, we identified student behaviour in the week leading up to the first summative assessment as a critical indicator of overall engagement throughout the year. These findings highlight opportunities for proactive interventions and support.

Our study provides a methodological contribution to EDM and LA research by refining an engagement measurement using VLE data and showcasing the effectiveness of Random Forest and GAMs in engagement analysis. Practically, our findings highlight critical periods for student engagement, such as early in the term and before summative assessments. These insights enable educators to identify students requiring additional support and inform strategic module planning, underscoring the potential of using VLE data proactively to improve teaching and student learning outcomes.

\bibliographystyle{apacite}


\end{document}